\documentclass[sigconf]{acmart}
\usepackage{xspace}
\usepackage{balance}
\AtBeginDocument{%
  }

\setcopyright{acmlicensed}
\copyrightyear{2025}
\acmYear{2025}
\setcopyright{cc}
\setcctype{by}
\acmConference[WWW Companion '25]{Companion Proceedings of the ACM Web
Conference 2025}{April 28-May 2, 2025}{Sydney, NSW, Australia}
\acmBooktitle{Companion Proceedings of the ACM Web Conference 2025 (WWW
Companion '25), April 28-May 2, 2025, Sydney, NSW, Australia}

\acmDOI{10.1145/3701716.3717853}
\acmISBN{979-8-4007-1331-6/2025/04}

\newcommand{\model}{Sketch-Search Agent\xspace}

\begin{document}

\title{Enhancing Product Search Interfaces with Sketch-Guided Diffusion and Language Agents}

\author{Edward Sun}
\email{edwardsun12895@g.ucla.edu}
\affiliation{%
  \institution{University of California, Los Angeles}
  \city{Los Angeles}
  \state{California}
  \country{USA}
}

\renewcommand{\shortauthors}{Edward Sun}

\begin{abstract}
The rapid progress in diffusion models, transformers, and language agents has unlocked new possibilities, yet their potential in user interfaces and commercial applications remains underexplored. We present \model, a novel framework that transforms the image search experience by integrating a multimodal language agent with freehand sketches as control signals for diffusion models. Using the T2I-Adapter, \model combines sketches and text prompts to generate high-quality query images, encoded via a CLIP image encoder for efficient matching against an image corpus. Unlike existing methods, \model requires minimal setup, no additional training, and excels in sketch-based image retrieval and natural language interactions. The multimodal agent enhances user experience by dynamically retaining preferences, ranking results, and refining queries for personalized recommendations. This interactive design empowers users to create sketches and receive tailored product suggestions, showcasing the potential of diffusion models in user-centric image retrieval. Experiments confirm \model's high accuracy in delivering relevant product search results.
\end{abstract}



\maketitle
\section{Introduction}
Sketches offer a simple and intuitive way to represent desired search images, tapping into humans' natural tendency to think visually \cite{Potter2013}. In the context of e-commerce, where billions of items are available, visual inputs like sketches can make searches more intuitive, efficient, and engaging. Recent advancements in image search methods have explored hand-drawn sketches using approaches such as triplet neural networks \cite{SketchWithStyle}, classical vision techniques \cite{MindFinder}, Linear Dynamical System models \cite{ContentRetrieve}, and Siamese CNNs with spatial feature pooling \cite{compositionalsketchsearch}, enabling searches based on simple outline features in a sketch.

However, current sketch-based image retrieval methods typically rely on a single search modality, using only sketches as input queries. Moreover, these methods often function solely as search engines, delivering direct results without offering personalization, guidance, or support throughout the search process—key factors for enhancing customer engagement on e-commerce platforms. Finally, existing methods lack interpretability and interactivity; the model's matching process is treated as a black box, providing results based on input sketches without transparency. This limits user engagement and requires them to carefully craft their queries in advance.

\begin{figure}
  \includegraphics[width=\columnwidth]{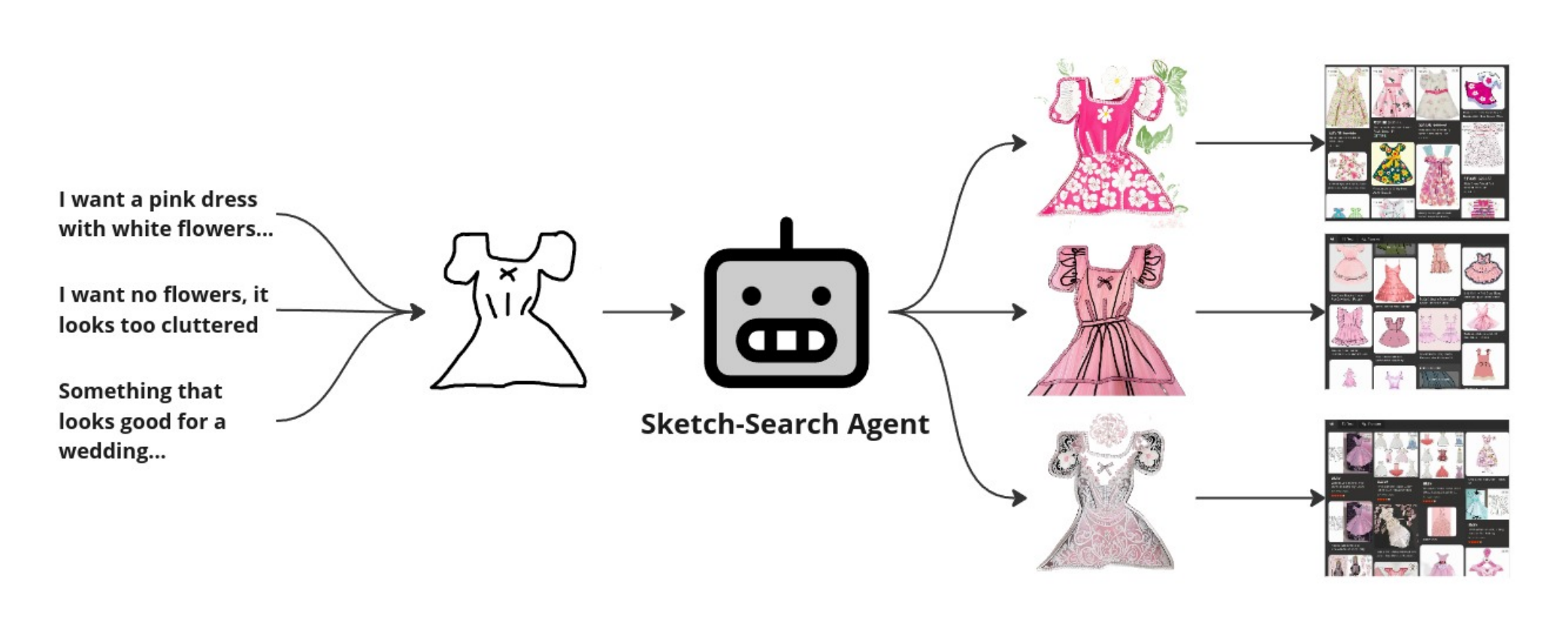}
  \caption{\model enables multimodal product search by integrating sketches with natural language refinement and personalized feedback from a language agent.}
  \Description{\model agent interaction example}
  \label{fig:interact}
  \vskip -0.20in 
\end{figure}

This paper presents a novel framework combining language agents with a memory module to retain user preferences and past queries, alongside a diffusion model enhanced by a T2I adapter \cite{T2I-Adapter}. The system refines input sketches into diffusion-generated image queries for reverse image search, dynamically optimizing textual conditions via API-connected language agents. It incorporates user preferences and product context to deliver personalized suggestions, visualizes ideal products, and enhances sketch-based searches through an intuitive chat interface as seen in Figure \ref{fig:interact}. Using CLIP embeddings for semantic alignment \cite{radford2021learningtransferablevisualmodels}, the framework supports reverse image search engines ranging from local models to Bing Image Search \cite{bing-search, CV4Image_Search, suma2024amesasymmetricmemoryefficientsimilarity}.

Our approach combines sketch and language modalities in a user-friendly manner, functioning as both a search tool and an e-commerce assistant. It remembers user preferences, engages in interactive dialogues, and allows users to explain their needs while interpreting the agent’s reasoning behind recommendations and image adjustments.

\section{Related Works}
\begin{figure*}[ht]
  \includegraphics[width=0.95\textwidth]{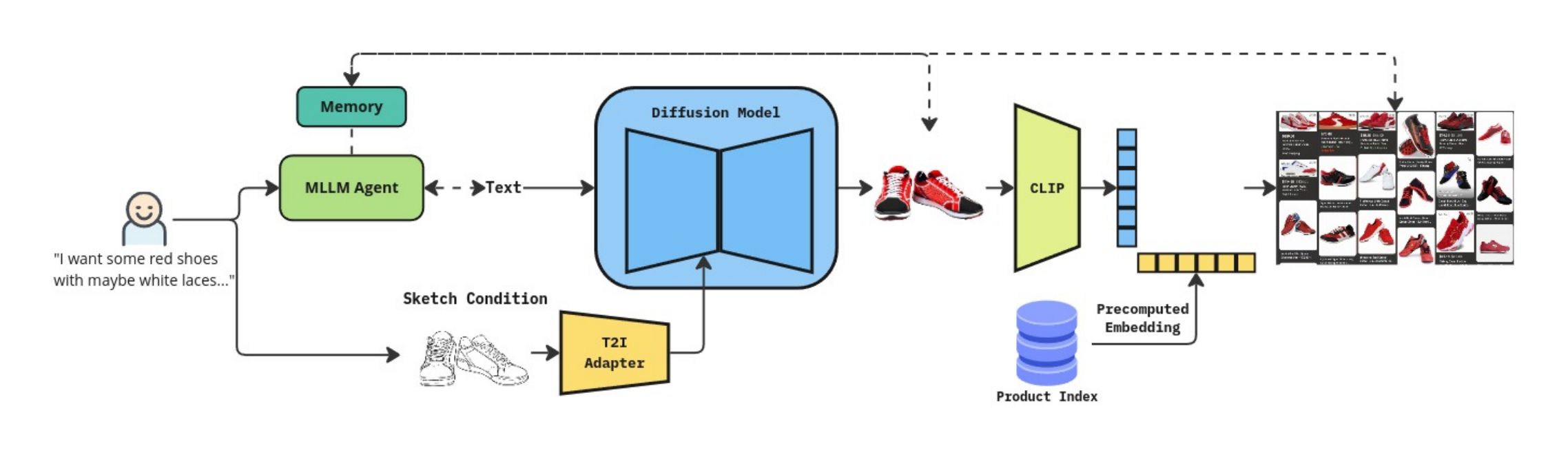}
  \caption{\model framework outline. Dashed lines represent possible information routes depending on the agent's choices and tool usage.}
  \Description{\model agent outline}
  \label{fig:outline}
\end{figure*}
\textbf{LLM Agents} Large Language Model (LLM) agents excel in tool usage \cite{huang2024planningcreationusagebenchmarking}, reasoning \cite{xiao2024logicvistamultimodalllmlogical}, and code generation \cite{jimenez2024swebenchlanguagemodelsresolve, wang2024scibenchevaluatingcollegelevelscientific}. They have been applied in high-frequency trading \cite{xiao2025tradingagentsmultiagentsllmfinancial}, biological discovery \cite{xiao2024proteingptmultimodalllmprotein, xiao2024rnagptmultimodalgenerativerna}, social interaction simulations \cite{zhao2024competeaiunderstandingcompetitiondynamics}, software engineering, robotic planning \cite{huang2023voxposercomposable3dvalue, liang2023codepolicieslanguagemodel}, and even as interactive interfaces to design 3D models \cite{lu2024enablinggenerativedesigntools}. In e-commerce, LLMs often act as chatbots with retrieval-augmented contexts or fine-tuned instructions \cite{peng2024ecellmgeneralizinglargelanguage}, and some explore API-based navigation \cite{chen2024chatshopinteractiveinformationseeking}. However, their role in visual and interactive search remains limited. We extend these applications with a framework incorporating visual generation, sketch-based search, personalization, and memory-enabled agents.

\textbf{Conditioned Diffusion Models} Advances in diffusion models utilize external conditions such as class \cite{ho2020denoisingdiffusionprobabilisticmodels, xing2024crossconditioneddiffusionmodelmedical}, text \cite{brooks2023instructpix2pixlearningfollowimage}, and image-based inputs \cite{T2I-Adapter, zhang2023addingconditionalcontroltexttoimage}, alongside guidance techniques like classifier-free methods \cite{ho2022classifierfreediffusionguidance, lin2024ctrlxcontrollingstructureappearance}. These innovations enable realistic, condition-guided image generation, presenting opportunities for sketch-to-image search in commercial applications with interpretable and interactive results.

\textbf{Sketch-Based Image Retrieval} Traditional sketch-based image retrieval focuses on efficient matching using techniques like triplet networks or query tensors for single-object models \cite{SketchWithStyle, SketchGuidedDiffusion, compositionalsketchsearch, SketchBasedImageRetrievalDomainLearning}. However, these methods lack interactivity, requiring users to craft precise queries using free-hand sketch only—limiting their practicality in e-commerce. To address this, we propose integrating LLM agents and diffusion models to enhance interactivity, providing recommendations, visualizing ideal products, and enabling an intuitive browsing experience.

\section{Methods}
\model\ is composed of two core components: a multimodal large language model (MLLM) agent and a T2I-Adapted diffusion model. In our experiments, we employed the pre-trained \verb|gpt-4o-mini| as the primary MLLM for its speed, paired with the OpenAI \verb|text-embedding-3-small| model for memory retrieval.

For image generation, we utilized a pretrained T2I adapter and \verb|stable-diffusion-xl| as the foundational model. To enhance generation speed and reduce \model's overall latency, we integrated Progressive Adversarial Diffusion Distillation, as described in SDXL-Lightning \cite{lin2024sdxllightningprogressiveadversarialdiffusion}. The inference process is conducted in two steps, along with the T2I adapter for rapid generation based on both sketch and text inputs.

\subsection{Module Definitions}
\textbf{MLLM Agent:} A Multimodal Large Language Model (MLLM) agent $A$ is defined with memory state $M_t$ and tool access $\tau$, based on the ReACT prompting scheme \cite{yao2023reactsynergizingreasoningacting}. The agent leverages tools to query a diffusion model, retrieve items, generate images, and perform other tasks. For brevity, we will highlight when a module functions as a tool without formally defining each one. 

At each step $t \in \{1, 2, \ldots\}$, the user provides a sketch $S_t$ and a natural language query $Q_t$ (e.g., “I want something for an upcoming wedding—preferably white with…”). The sketch $S_t$ remains unchanged if the user modifies $Q_t$ for refinement.

The agent processes input through one or more reasoning/tool usage rounds, either responding directly (e.g., suggestions or answers) or transforming $Q_t$ into a refined query $Q_t'$ for the diffusion model. This transformation, guided by few-shot examples, optimizes conditioning by removing extraneous details while preserving user intent. The agent is defined as:
\begin{equation}
    A(Q_t, M_t) = 
    \begin{cases} 
    R_t & \text{if an immediate response is required}, \\
    Q_t' & \text{if refining the sketch condition}.
    \end{cases}
\end{equation}

\textbf{T2I Adapted Diffusion} We employ a T2I-adapted diffusion model $G$, which takes a sketch image $S$ and conditioning text $Q'$ as inputs. The model generates an image $I'$ based on these inputs, ensuring alignment with both conditionings. Formally, the T2I-adapted diffusion model is defined as:
\begin{equation}
    G(S, Q') = I'
\end{equation}

This module is implemented as a tool available to the agent in our code. Similarly, we have another tool that enables the agent to retrieve the generated image in order for the agent to generate suggestions of changes/queries or use that to inform user preferences.

\textbf{Image Search} We designed \model such that any image search method can be used at this step from off the shelf methods like Bing Search \cite{bing-search} to other local methods of searching indices. Here for our experiments, we utilize a $CLIP$ encoder to generate our embeddings, which maps an image $I$ to a d-dimensional vector: 
\begin{equation}
CLIP(I) \in \mathbb{R}^d
\end{equation}
We search over a catalog of $N$ products locally, consisting of images $\{I_1, I_2, \ldots, I_N\}$. We pre-compute embeddings for these $N$ images by encoding each $I_k$ as $e_k = CLIP(I_k) \in \mathbb{R}^d$. Additionally, we implement tools for the MLLM agent to retrieve the list of search results and their images in order for the agent to sort by relevance and suggest products directly.

\subsection{User-Agent Interaction Steps}
As shown in Figure \ref{fig:outline}, a single interaction step offers multiple possibilities for the agent, such as calling tools to retrieve context, providing user suggestions and recommendations based on search results and memory, or answering queries. Here, we focus on the search process as it is the framework's core and most frequently used function. Given user input $(S_t, Q_t)$, where $Q_t$ is a search query to condition the image, \model executes the following steps:
\begin{enumerate}
    \item \textbf{Condition Generation:} $A(\cdot)$ generates a condition $Q_t'$ for $G(\cdot)$ based on the user query.
    \item \textbf{Image Generation:} The agent calls tools to invoke the diffusion model, producing the conditioned image $I_t' = G(S_t, Q_t')$.
    \item \textbf{Embedding and Ranking:} The generated image $I_t'$ is embedded via $e_t' = CLIP(I_t')$, where $e_t' \in \mathbb{R}^d$. Precomputed product embeddings $e_k = CLIP(I_k)$ for all $I_k \in D$ are used with cosine similarity $sim(e_t', e_k)$ to rank $N$ items, yielding the retrieved product list $\Pi_t$:
\begin{equation}
    \Pi_t = \arg\max_{k \in \{1,\ldots,N\}} \mathrm{sim}(e_t', e_k).
\end{equation}
The ranked list $\Pi_t$ is presented to the user. If required, the agent may also call tools to retrieve this list or product additions $R_t$.

Note here that the method of searching is easily swappable for any other image search method such as Bing Search \cite{bing-search} or other content-based image retrieval techniques \cite{DeepInstance}.
\end{enumerate}

In the next interaction $t+1$, the user provides a new query $Q_{t+1}$, which serves as feedback $F_t$ for $Q_t$ to update the memory state and build personalization. The memory is updated using the function $U(\cdot)$:
\begin{equation}
    M_{t+1} = U(M_t, F_t, \Pi_t).
\end{equation}
The update function \( U(\cdot) \) is implemented by embedding documents using OpenAI's \texttt{text-embedding-3-small} model, with the resulting embeddings stored in a vector database for efficient retrieval.

\section{Experiments}
To evaluate the preliminary effectiveness of \model, we designed two experiments. The first focused on measuring the average latency of generation and search processes. The second, more critical experiment assessed the success rate of searches, the relevance of product matches, and the helpfulness of user recommendations.

\textbf{Data} To adapt \model for e-commerce, we used a real-world dataset, the Google Product Image Embedding dataset \cite{zhu2024marqoecommembed_2024}, extracting a 10K subset as our product index. Each data point includes dataset-provided labels and MLLM-generated natural language tags. We also created 400 sketch-condition inputs, comprising 200 human-drawn sketches and 200 edge-detected images processed with Pixel Difference Networks (PidiNet) \cite{pdc-PAMI2023, su2021pdc, su2019bird} to scale our dataset better. To simulate dynamic scenarios, an external judge model, \verb|Gemini-Flash-2.0|, evaluated inputs based on text conditions. This model, chosen for its superior vision understanding, ensured impartiality by avoiding overlap with \model's backbone MLLM to avoid LLM preferences for their own generations \cite{panickssery2024llmevaluatorsrecognizefavor}.

\textbf{Models} In our experiments, we compare \model to its ablations: (1) \verb|No-Refine|, where the user query is directly fed to the diffusion model without refinement; (2) \verb|Tools-Only|, where the agent has no memory and relies solely on tools; (3) \verb|Memory-Only|, where the agent uses memory but no tools, except the diffusion module necessary for searches; and (4) the full \model, as outlined in Figure \ref{fig:outline}.

\textbf{Latency} We measured the average latency of \model in its most demanding scenario: generating an image and performing a search. This involves the combined use of the MLLM agent, diffusion modules, and reverse image search. To evaluate performance, we ran 400 experiments on our 10K index dataset using a 400-image dataset and averaged the results.

\textbf{Success Rate} We evaluate \model's success rate as the ratio of correct matches to total samples by aligning sketch queries and text conditions with ground truth labels. For each of the 400 samples, a judge model inputs a text condition and corresponding sketch into \model. To retain structural details often lost in diffusion-generated images, both the sketch and text condition are uploaded together. The judge then compares the simulated user intent (sketch from the dataset and LLM generated text condition) to the dataset's natural language tags, assigning a binary accuracy label. The success rate is the average accuracy across all 400 samples.

\textbf{Personalization Score} We assess \model's personalization using its memory modules, preference ranking, and chat-based mechanisms. Before evaluation, we pre-load the memory module with user preferences to simulate multi-round memory of \model. A separate LLM judge, \verb|Gemini-Flash-2.0|, is provided with the user's context, preferences, use case, generated image, and search rankings to evaluate personalization on a 5-point Likert scale. The results are averaged over our 400 samples.

\section{Results}
Here we present the results of our evaluation of \model through the ablations \verb|No-Refine|, \verb|Tools-Only|, \verb|Memory-Only|, and \model.

\textbf {Latency} We present the latency analysis of \model, as detailed in Table \ref{tab:latency}. Each major generative module's average latency was measured over 400 search queries, focusing on the bottleneck: worst-case scenarios requiring tool calls, diffusion model generations, and CLIP embeddings for search. The primary contributor to latency is the diffusion model, with a maximum runtime of approximately 2.3 seconds. This is mitigated by our use of a distilled diffusion model (2-step generation) and the faster \verb|gpt-4o-mini|, ensuring responsiveness and efficiency. Overall, \model achieves an average worst-case latency of 4 seconds, a reasonable upper bound given the system's complexity with two generative models.

\begin{table}[h!]
\centering
\begin{tabular}{ c|c c c }
\hline
\textbf{Method} & \textbf{Min} $\downarrow$ & \textbf{Max} $\downarrow$ & \textbf{Mean} $\downarrow$ \\ \hline
\textbf{Distilled \& T2I Diffusion} & 1.74 & 2.32 &  1.92 \\ 
\textbf{Search} & 0.18 & 0.23 &  0.21 \\ 
\textbf{MLLM} & 0.92 & 1.44 &  1.11 \\ 
\textbf{\model} & 3.24 & 4.45 & 3.86 \\ \hline
\end{tabular}
\caption{Latency of each step in \model over a 10K real commercial products dataset (measured in seconds).}
\label{tab:latency}
\end{table}

\begin{figure}[ht]
  \includegraphics[width=\columnwidth]{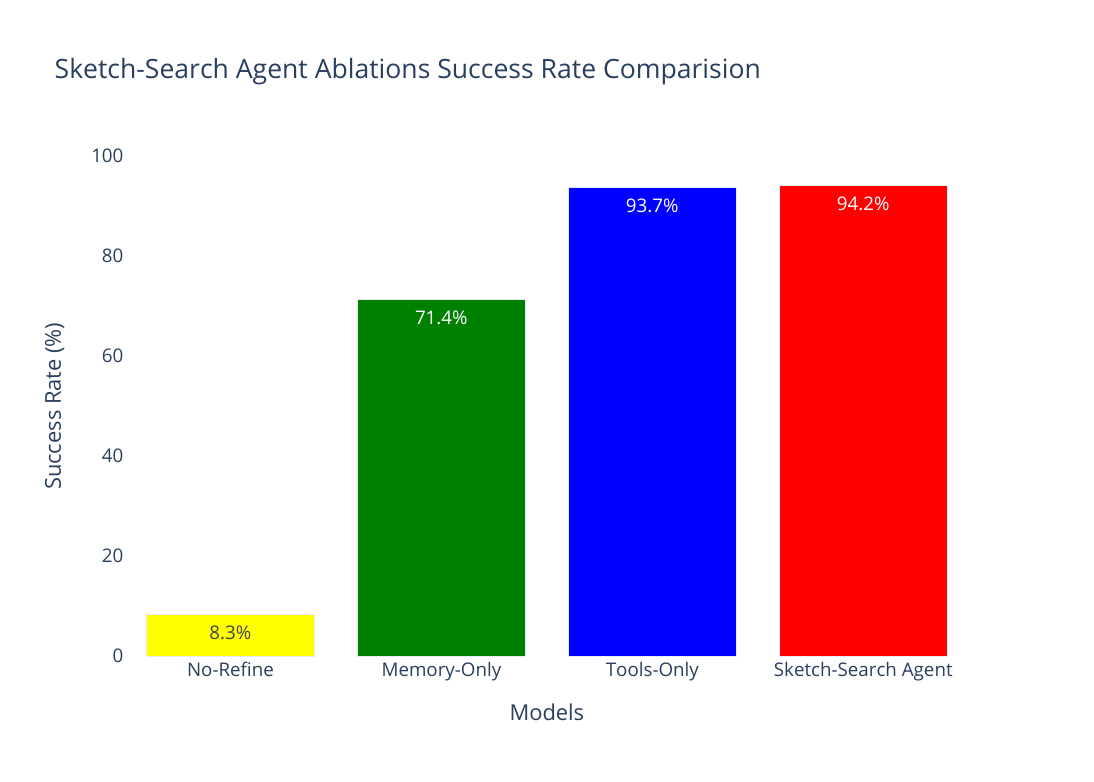}
  \caption{Averaged success rate ablations of \textbf{\textcolor{red}{\model}}}
  \Description{\model success rate results}
  \label{fig:success}
\end{figure}

\textbf{Success Rate} Figure \ref{fig:success} shows the average success rate across 400 queries, evaluated by an MLLM judge. The full \model outperforms its ablations, highlighting the importance of key components. Without prompt-to-text conditioning refinement, success rates drop sharply as the diffusion model struggles to process language, leading to poor image generation and inaccurate searches. Removing tool calls also reduces performance, as the model loses access to critical contextual information.

Interestingly, the tool-only variant performs comparably to the full \model in single-round search tasks, which don’t require interaction history. Combined with latency results, this demonstrates \model’s efficiency and versatility. By integrating sketches, language, and contextual reasoning, \model enables fast, creative, and accurate e-commerce product searches, making it a powerful tool for enhancing user experience.

\begin{figure}[ht]
  \includegraphics[width=\columnwidth]{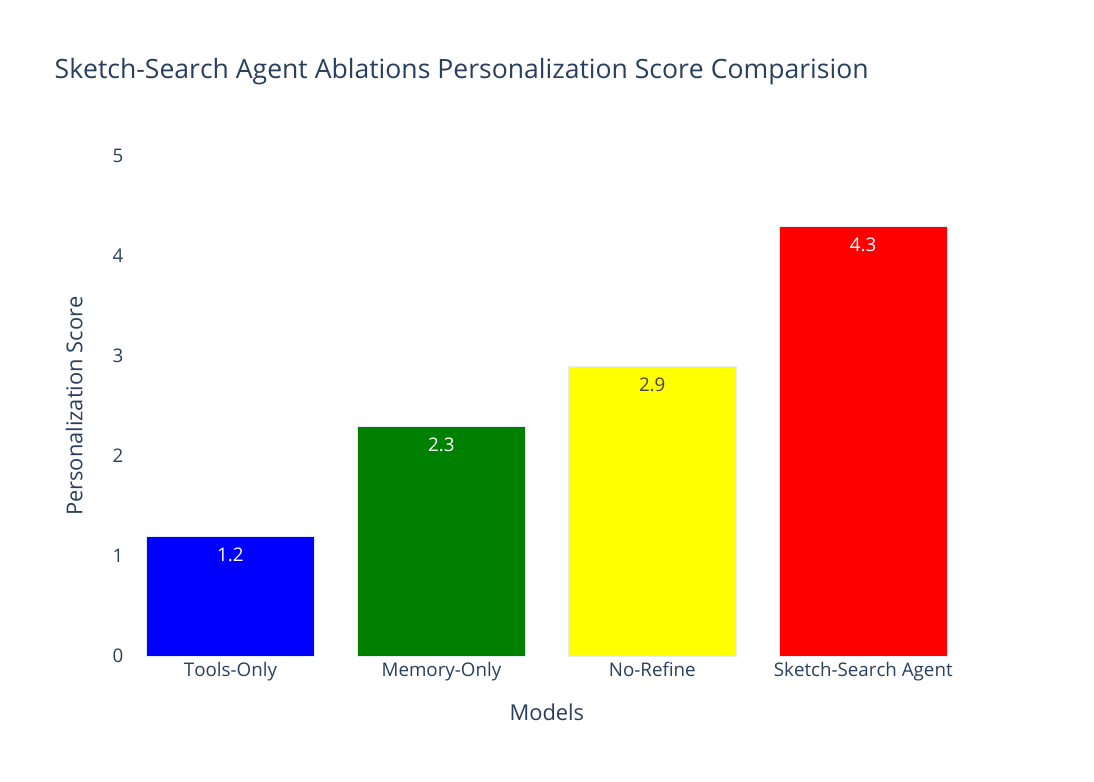}
  \caption{Averaged personalization score ablations of \textbf{\textcolor{red}{\model}}}
  \Description{\model personalization score results}
  \label{fig:personal}
\end{figure}

\textbf{Personalization Score} Figure \ref{fig:personal} presents the personalization measurements assessed by our MLLM judge, averaged across all 400 samples. As shown, the "tools-only" ablation performs poorly on this metric, primarily because, without a memory module, it cannot effectively retain user preferences and thus fails to deliver personalized results. Its low score stems from the few instances where it might guess personalization correctly based solely on the query and sketch context.

Interestingly, while the "memory-only" ablation performs slightly better, its performance remains poor. This is likely because, without tools, the agent cannot reference the generated images or their rankings to deliver detailed personalizations. Finally, the absence of prompt refinement produces issues similar to those in the success rate experiment: the generated images lack relevance to the query, as the diffusion conditions are framed in dialogue-like language. These findings validate the design choices in \model, highlighting the importance of using both tools to retrieve user preferences and results, integrating a multi-round agent memory, and refining prompts to achieve effective personalization in sketch agents.

\section{Conclusion}
In this work, we present \model, a novel MLLM agent framework that utilizes diffusion models to conduct sketch-based image search for commercial applications. \model showcases the transformative potential of integrating diffusion models, multimodal agents, and sketch-based image search in a unified framework, redefining e-commerce and interactive search by addressing limitations like single modality reliance, minimal personalization, and low interactivity. Experimental results highlight its ability to deliver accurate, personalized recommendations with low latency, enhanced by memory modules, refinement mechanisms, and interactive dialogue. By leveraging multimodal inputs and user feedback, \model advances human-AI collaboration and lays a foundation for future innovations, including expanding our work with broader applications, faster diffusion inference like shortcut models, adaptive life-long learning, alternative guidance methods, and multi-agent frameworks to improve scalability, personalization, and efficiency.

\bibliographystyle{ACM-Reference-Format}
\balance
\bibliography{sample-base, software}

\appendix
\end{document}